\documentclass[twocolumn,prl,aps,amsmath,amssymb,floatfix,showpacs,longbibliography,notitlepage,superscriptaddress]{revtex4-1}

\usepackage[usenames,dvipsnames]{xcolor}
\definecolor{af}{RGB}{255, 127, 14}
\definecolor{fm}{RGB}{31, 119, 180}

\usepackage{graphicx}
\graphicspath{{./figs/}}
\usepackage{bm}

\usepackage[colorlinks=true,linkcolor=blue,citecolor=blue,urlcolor=blue]{hyperref}

\usepackage{tabularx}
\newcolumntype{C}[1]{>{\centering\arraybackslash}m{#1}}
\newcolumntype{L}[1]{>{\raggedright\arraybackslash}m{#1}}
\usepackage{multirow}

\newcommand\3{$_3$}
\newcommand\rtriangle{${\color{af}\triangleright}$}
\newcommand\ltriangle{${\color{af}\triangleleft}$}

\newcommand\geneva{Department of Quantum Matter Physics, University of Geneva, CH-1211 Geneva, Switzerland}
\newcommand{\unimore}{Dipartimento di Scienze Fisiche, Informatiche e Matematiche, University of Modena and Reggio Emilia, IT-41125 Modena, Italy}

\AtBeginDocument{\usepackage{booktabs}}
\begin{document}
\title{Magnetism and  stability of all primitive stacking patterns in bilayer chromium trihalides}

\author{Marco Gibertini}
\affiliation{\unimore}
\affiliation{\geneva}

\date{\today}

\begin{abstract}
Chromium trihalides, CrX\3 (with X = Cl, Br, I), are a family of layered magnetic materials that can be easily exfoliated to provide ferromagnetic monolayers. When two layers are stacked together to form a bilayer the interlayer exchange coupling can be either ferromagnetic or antiferromagnetic depending on the stacking sequence. Here we combine crystallographic arguments based on the close-packing condition with first-principles simulations to enumerate all possible stacking patterns in CrX\3 bilayers that preserve the spatial periodicity of each layer. We recover all configurations observed in bulk crystals and disclose stacking sequences with  no bulk counterpart where the two layers have opposite \emph{chirality}. Stacking sequences are ranked according to their relative stability and a preferential interlayer magnetic ordering is assigned to each of them. Simulations provide a consistent picture to frame all current experimental observations on bulk and exfoliated CrX\3 crystals, with interesting implications for future measurements, including synthetic bilayers with non-standard stacking patterns. 
\end{abstract}

\maketitle

\section{Introduction}

Van der Waals magnetic materials have always been a source of fascinating phenomena~\cite{deJong1990} that are now attracting revived interest owing to the possibility of exfoliating these compounds down to the monolayer limit~\cite{Burch_review_2018,Gong_review_2019,Gibertini_review_2019,Mak_review_2019}.  Among these, the family of chromium trihalides (CrX\3, X = Cl, Br, I) has emerged as a promising playground for experimental~\cite{Huang2017,Song2018,Klein2018,Wang2018,Kim2018,Ghazaryan2018,Cai2019,Klein2019,Wang2019,Kim2019,Kim2019b,Kim2019hall} and theoretical~\cite{Sivadas2018,Soriano2019,Jiang2019theory,Jang2019,Lei2019,Cardoso2018,Tong2018,Pizzochero2020,Soriano2020} explorations. In particular, the magnetic properties of atomically thin CrI\3~\cite{Huang2017,Song2018,Klein2018,Wang2018,Kim2018} can be easily manipulated by external controls such as electric fields~\cite{Jiang2018field}, doping~\cite{Huang2018,Jiang2018doping}, and pressure~\cite{Song2019pressure,Li2019}, allowing the realization of spin-sensitive devices~\cite{Jiang2019,Song2019}, although practical applications are limited by the low critical temperature. 

Bulk CrX\3 crystals are magnetic insulators where spins are ferromagnetically aligned within each layer, while the interlayer exchange coupling depends on the halogen. In CrBr\3 and CrI\3, all layers share the same spin orientation, giving rise to an overall ferromagnetic (FM) behaviour~\cite{Tsubokawa1960,McGuire2015}, while CrCl\3 is a layered antiferromagnet with neighbouring layers having spins pointing in opposite directions (so called ``A-type" ordering)~\cite{Cable1961,McGuire2017}. Another difference arises from the lighter atomic number of Cl atoms, which leads to reduced spin-orbit coupling effects, so that spins are oriented parallel to the layers in CrCl\3 and perpendicular in CrBr\3 and CrI\3. 

In addition to the magnetic phase transition at low temperature, CrX\3 crystals undergo a structural phase transition at higher temperatures from a high-temperature monoclinic phase to a low-temperature rhombohedral phase~\cite{Morosin1964,McGuire2015,Djurdjic2018,Samuelsen1971}. Across the transition, the structure of each layer remains essentially unaffected, with the main difference between the two phases being the stacking sequence of the layers~\cite{McGuire2015}. At room temperature, only CrBr\3 is already in the low temperature structure, while CrI\3 and CrCl\3 are still in the monoclinic phase. At temperatures below the magnetic transition, instead,  all members of the chromium trihalide family share the same rhombohedral structure (see table~\ref{tab:exp}). 

When thinned down to few atomic layers, CrI\3 has been reported to display A-type antiferromagnetic (AF) order~\cite{Huang2017}, giving rise to a strong spin-filtering effect and a large magnetoresistance in tunnelling devices~\cite{Song2018,Klein2018,Wang2018,Kim2018}. The different magnetic ordering of atomically thin and bulk CrI\3 has been long puzzling, especially in view of the apparently similar behaviour of CrBr\3 and CrCl\3 multilayers with respect to their 3D counterparts~\cite{Ghazaryan2018,Kim2019,Kim2019hall,Klein2019,Wang2019,Cai2019}. First-principles simulations have provided a possible solution to this conundrum by demonstrating a strong connection between the stacking pattern and the magnetic ground state~\cite{Wang2018,Sivadas2018,Soriano2019,Jiang2019theory,Jang2019,Lei2019}, predicting that for CrI\3 the monoclinic phase should be AF while the rhombohedral phase is FM. 

Raman spectroscopy~\cite{Ubrig2019} and non-linear optical measurements~\cite{Sun2019} have validated this picture by showing that bulk CrI\3 is rhombohedral at low temperature, consistently with the bulk FM ordering, while in atomically-thin samples the stacking sequence is monoclinic, thus explaining the AF interlayer coupling (also present in superficial layers~\cite{Niu2020,Li2020}). A monoclinic structure has been reported also in thin CrCl\3, while keeping the same A-type AF order as in its bulk rhombohedral form, although with an enhanced interlayer coupling~\cite{Klein2019}. For CrBr\3, no experimental result is currently available on the stacking pattern in thin crystals, although clear indications exist that they preserve the bulk FM interlayer coupling~\cite{Ghazaryan2018,Kim2019,Kim2019hall}.

To close the circle and provide a consistent picture for all current experimental results on atomically thin CrX\3 samples, here we combine general crystallographic arguments with first-principles simulations to explore the magnetic ground state and the relative stability of all possible stacking patterns in bilayer chromium trihalides that  preserve the spatial periodicity of each layer (i.e.\ primitive).  We extend previous results on CrI\3 to CrBr\3 and CrCl\3 and disclose stacking sequences that have no bulk counterpart but could be relevant in synthetic crystals~\cite{Chen2019}. The theoretical scenario that arises suggests that  atomically thin CrBr\3 should display a rhombohedral structure, differently from the other chromium trihalides CrI\3 and CrCl\3, as a result of the higher critical temperature for the structural phase transition, which makes CrBr\3 rhombohedral --and not monoclinic-- at the temperatures at which exfoliation takes place.

\begin{table}[t]
\caption {Low-temperature magnetic order (ferromagnetic, FM, or antiferromagnetic, AF) and stacking symmetry (rhombohedral or monoclinic) in bulk and multilayer chromium trihalides CrX\3 as experimentally reported in the literature.
\label{tab:exp}}
\vspace{2mm}
{\setcitestyle{super}
\begin{tabularx}{\linewidth}{l >{\centering\arraybackslash}X
>{\centering\arraybackslash}X >{\centering\arraybackslash}X }
\toprule
 & CrCl\3 & CrBr\3 & CrI\3 \\
\cmidrule{2-4}
\multirow{2}{*}{bulk} & AF\cite{Cable1961,McGuire2017}  &  FM\cite{Tsubokawa1960} & FM\cite{McGuire2015}  \\
                                 & rhomb.\cite{Morosin1964} &  rhomb.\cite{Samuelsen1971} & rhomb.\cite{McGuire2015,Djurdjic2018} \\
\midrule
\multirow{2}{*}{multilayers} &AF\cite{Cai2019,Wang2019,Klein2019,Kim2019} & FM\cite{Ghazaryan2018,Kim2019hall,Kim2019}  &   AF\cite{Huang2017,Song2018,Klein2018,Wang2018,Kim2018} \\
& monocl.\cite{Klein2019} &  -- &   monocl.\cite{Sun2019,Ubrig2019}\\
\bottomrule
\end{tabularx}}
\end {table}

\section{Methods}
First-principles simulations are performed within  density-functional theory (DFT)  using the Quantum ESPRESSO distribution~\cite{Giannozzi2009,Giannozzi2017}. To account for van der Waals interactions between the layers, the spin-polarised extension~\cite{Thonhauser2015} of the revised  vdw-DF2 exchange-correlation functional~\cite{Hamada2014,Lee2010} is  adopted, truncating spurious interactions between artificial periodic replicas along the vertical direction~\cite{Rozzi2006,Ismail-Beigi2006,Sohier2017}. The Brillouin zone is sampled with a $8\times8\times1$ $\Gamma$-centered Monkhorst-Pack grid. Pseudopotentials are taken from the Standard Solid-State Pseudopotential (SSSP) accuracy library~\cite{Prandini2018,Garrity2014,DalCorso2014} (v1.0) with increased cutoffs of 60 Ry and 480 Ry for wave functions and density, respectively. Total energy calculations as a function of the relative displacement between the layers are performed without atomic relaxations, by taking the structure of DFT-relaxed monolayers with the experimental lattice parameter and interlayer separation. For line scans, atomic positions are relaxed by reducing the force acting on atoms below a threshold of 26~meV/\AA, while keeping fixed the in-plane coordinates of Cr atoms.  The threshold is reduced to 3~meV/\AA\ without constraints to evaluate the relative stability of the different stacking configurations, while also optimising  the lattice parameter and cell angle (for non-hexagonal systems) using an equation-of-state approach. Calculations are managed and automated using the AiiDA materials informatics infrastructure~\cite{Pizzi2016,Huber2020}.

\section{Results and discussion}

\begin{figure}
\includegraphics{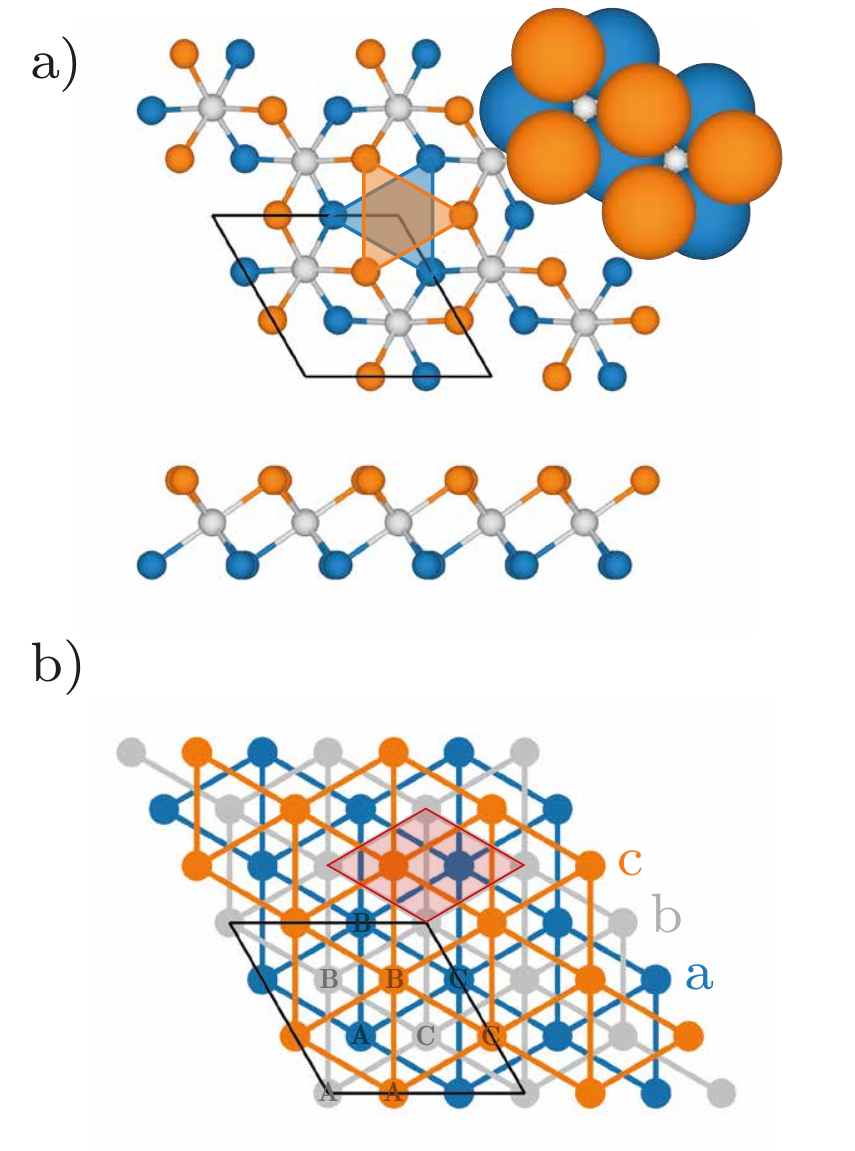}
\caption{Panel a): Top and later views of the crystal structure of monolayer CrX\3, with a black solid line showing the perimeter of the unit cell. Chromium atoms are reported in grey, while halogens are shown with different colours (orange and blue) to distinguish the planes above and below the layer of Cr atoms. In the top right corner the size of halogen atoms is exaggerated to emphasise their close-packed arrangement. Cr atoms occupy 2/3 of the octahedral interstitials, with a single empty octahedron per unit cell. The vertices of the empty octahedron are highlighted with orange and blue shaded triangles in the top and bottom halogen plane, respectively. Panel b): Three possible triangular sublattices (grey, orange, and blue) that are consistent with a close-packed arrangement. The primitive unit cell of the triangular sublattices is highlighted with a red shade, while the overall unit cell of CrX\3 is shown in black (and corresponds to a $\sqrt{3}\times\sqrt{3}$ supercell rotated by 30$^\circ$). \label{fig:mono}}
\end{figure}

The crystal structure of monolayer CrX\3 is reported in Fig.~\ref{fig:mono}a), and consists of three atomic planes: a layer of chromium atoms (grey) sandwiched between halogen layers, reported with different colours (orange and blue) to distinguish the top and bottom plane. The structure can be rationalised by noting that each halogen layer forms a planar triangular sublattice (although slightly distorted) and that the two planes are close-packed (see the top right corner of Fig.~\ref{fig:mono}a). Chromium atoms occupy octahedral interstitials, which themselves form a third triangular lattice, although only 2/3 of the sites are occupied. The three close-packed triangular sublattices are shown in Fig.~\ref{fig:mono}b) with different colours (blue, grey, and orange)  and named a, b, and c. Because 1/3 of the octahedral interstitials are empty, the unit cell of CrX\3 (black solid line) is larger than the unit cell of a single triangular sublattice (red shade), and represents a $\sqrt{3}\times\sqrt{3}$ supercell rotated by 30$^\circ$. 

Once the halogen triangular sublattices are specified, the Cr sublattice is enforced by the close-packing condition and the only degree of freedom is the choice of the empty site among  three possibilities  (marked A, B, and C, in Fig.~\ref{fig:mono}). Thus the structure of a monolayer can be identified by specifying the two halogen triangular sublattices and the empty site in the chromium layer. For instance, in Fig.~\ref{fig:mono}a) the bottom halogen plane corresponds to the ``a" sublattice, while the top halogen plane to the ``c" sublattice. Cr atoms belong to the ``b" sublattice with the ``A" site empty. We denote this structure as ``aA$_{\rm b}$c", where the left (right) small letter denotes the bottom (top) halogen sublattice while the capital letter identifies the empty site in the Cr layer, with a subindex specifying the corresponding sublattice. 

We can thus see that CrX\3 monolayers can exist in two inequivalent forms. Fixing an arbitrary choice of the origin at the empty site uniquely determines the Cr layer to be A$_{\rm b}$. The close-packing condition allows only two possible choices for the halogen planes: aA$_{\rm b}$c (as in Fig.~\ref{fig:mono}a) and cA$_{\rm b}$a (with top and bottom halogen planes exchanged). The difference between the two can be best visualised by considering the halogen atoms forming the octahedral cage around the empty site. In the aA$_{\rm b}$c case of Fig.~\ref{fig:mono}a), the top (bottom) halogens make a right (left) pointing triangle, while the opposite is true in the cA$_{\rm b}$a case. We thus have two possibile \emph{chiralities} for a monolayer, identified by the direction of the triangles in the top plane. The left chirality arises when the labels of the sublattices in the three atomic planes are an even permutation of ``abc" (as in Fig.~\ref{fig:mono}a), while the right chirality occurs for odd permutations (e.g.\ ``cba"). The two chiral structures can be obtained one from the other either by exchanging the top and bottom halogen planes, or equivalently by a   $60^\circ$ rotation or a mirror reflection (as it is typical of enantiomers).

\begin{table}[t]
\caption {Possible primitive stacking patterns in bilayer CrX\3 that satisfy the close-packing condition. Assuming the first layer to be in the aA$_{\rm b}$c configuration (see text for details), the second layer can either display the same or opposite chirality. In each case, the configuration of the second layer is reported, together with the in-plane component of the relative translation between the layers. Inequivalent stacking patterns are given short names for simplicity and associated with the corresponding point group, reported in both the International and Sch\"onflies (in parenthesis) notation.
\label{tab:configurations}}
\vspace{2mm}
\begin{tabularx}{\linewidth}{L{1.1cm} L{1.3cm} L{1.1cm}  X L{1cm} L{1.4cm}}
\toprule
First layer& Chirality & Second Layer &  Translation &Short name & Point group  \\
\midrule
\multirow{14}{*}{aA$_{\rm b}$c \rtriangle}  & \multirow{6}{*}{same}
   & aA$_{\rm b}$c \rtriangle  & (0,0) & AA &  $\bar3$m ($D_{3d}$)\\
\cmidrule{3-6}
& & aB$_{\rm b}$c \rtriangle &  $(0,a/\sqrt{3})$ &AB  & \multirow{2}{*}{$\bar3$ ($S_6$)}  \\
& & aC$_{\rm b}$c  \rtriangle&  $(a,a/\sqrt{3})/2$ &AC &  \\
\cmidrule{3-6}
& & bA$_{\rm c}$a  \rtriangle&  $(a/3,0)$ &\multirow{3}{*}{HT} & \multirow{3}{*}{$\rm2/m$ ($C_{2h}$)} \\
& & bB$_{\rm c}$a  \rtriangle&  $(a,a\sqrt{3})/3$  &&  \\
& & bC$_{\rm c}$a  \rtriangle&  $(-a,a\sqrt{3})/6$ &&  \\
\cmidrule{2-6}
& \multirow{6}{*}{opposite} & aA$_{\rm c}$b  \ltriangle&  $(a/3,0)$  & \multirow{3}{*}{rHT} & \multirow{3}{*}{m ($C_s$)} \\
& & aB$_{\rm c}$b  \ltriangle & $(a,a\sqrt{3})/3$ &  &  \\
& & aC$_{\rm c}$b  \ltriangle&  $(-a,a\sqrt{3})/6$ &  &  \\
\cmidrule{3-6}
& & bA$_{\rm a}$c  \ltriangle&  $(a,a\sqrt{3})/6$ & \multirow{3}{*}{rHT'} & \multirow{3}{*}{m ($C_s$)}  \\
& & bB$_{\rm a}$c  \ltriangle&   $(2a/3,0)$ & &  \\
& & bC$_{\rm a}$c  \ltriangle&  $(-a,a\sqrt{3})/3$& &  \\
\bottomrule
\end{tabularx}
\end {table}

We now turn our attention to the possible configurations of bilayer CrX\3. We restrict to \emph{primitive} stacking arrangements that preserve the translational invariance of each layer, that is with the same (primitive) unit cell as a monolayer.  The most stable configurations of the bilayer are expected to follow the same close-packing conditions of the monolayer. Choosing for definiteness the bottom layer in the aA$_{\rm b}$c form, the bottom halogen plane of the second layer needs to be in the ``a" or ``b" sublattice. In each case, we have then two possible choices for the top halogen plane sublattice that are compatible with the close-packing condition, and three possibilities for the empty site of the Cr layer. We thus expect $2\times2\times3=12$ possible stable configurations of the bilayer, listed in table~\ref{tab:configurations}. In half of them the two layers share the same chirality, while the chirality is different in the other six cases. 

\begin{figure}
\includegraphics[width=\linewidth]{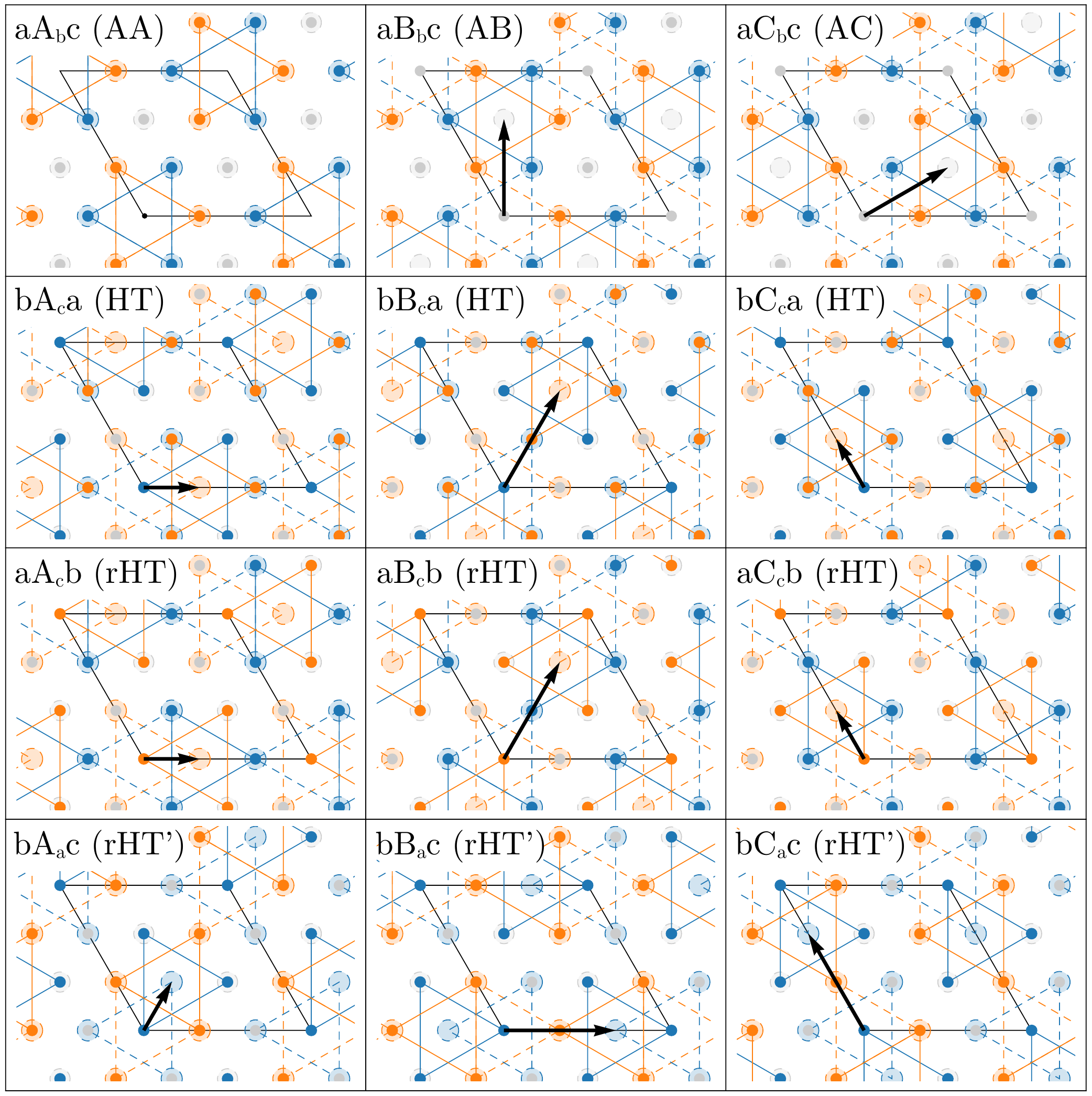}
\caption{Graphical representation of the primitive stacking patterns listed in table~\ref{tab:configurations}. Chromium atoms are shown in grey, while halogens above and below the Cr sheet are reported in orange and blue, respectively. The first (bottom) layer is assumed to be in the aA$_{\rm b}$c configuration (see text for details) and is shown with bigger, shaded symbols with dashed contours. The second (top) layer takes the configuration specified in the upper left corner, with the corresponding short name in parenthesis, and is shown with full symbols.  Dashed (solid) orange and blue triangles highlight the configuration of the top and bottom halogen atoms around the empty site in the Cr sublattice of the first (second) layer. The black solid line marks the primitive unit cell contour, while thick arrows denote the in-plane component of the relative translation between the layers (folded inside the unit cell). \label{fig:conf}}
\end{figure}

In Fig.~\ref{fig:conf} we report a schematic picture of the different stacking sequences in table~\ref{tab:configurations}, where  the orange and blue dashed (solid) lines highlight the triangles corresponding respectively to the top and bottom halogens around the empty site in the first (second) layer. When the chirality is the same, we can interpret the second layer as obtained from the first one by a rigid translation, which has both an in-plane (shown in the figure) and out-of-plane component. When the chirality is different, the second layer can still be obtained from the fist one by a rigid translation, but we need to perform first a rotation by $60^\circ$ (or a vertical mirror reflection). 

In the bulk form, all layers have the same chirality, so we should expect to find the bulk stacking sequences among these cases. Indeed, two consecutive layers in the rhombohedral phase share the same sublattices in the halogen planes, while the position of the empty site is different. In table~\ref{tab:configurations}, this situation corresponds to the second layer being either aB$_{\rm b}$c or aC$_{\rm b}$c (with the first one being aA$_{\rm b}$c). Since the information on the halogens is redundant in the two layers, these configurations are typically denoted simply as AB or AC from the position of the empty site, and they are equivalent up to a redefinition of the lattice vectors (and thus have the same point group). In principle, also another possibility arises when the two layers share the same halogen sublattices, that is when also the empty position is the same, thus corresponding to an AA stacking sequence. The high-temperature (HT) monoclinic stacking arrangement can also be recovered. It corresponds to the second layer being bA$_{\rm c}$a, bB$_{\rm c}$a, or bC$_{\rm c}$a, corresponding to a rigid translation along one of three equivalent high-symmetry directions. Translations along the same lines but in opposite directions would give rise to a HT' configuration (with the second layer  being either cA$_{\rm a}$b, cB$_{\rm a}$b, or cC$_{\rm a}$b). This HT' configuration is indistinguishable from the standard HT bulk arrangement when only Cr atoms are considered.   Nonetheless, it does not satisfy the close-packing condition as the halogen planes facing the van der Waals gap share the same ``c" sublattice and thus sit on top of each other, so that this configuration is expected to be unstable.

Stacking sequences where the two layers have different chirality do not have a bulk counterpart, although they can be realised in synthetic bilayers grown by molecular beam epitaxy~\cite{Chen2019}. Among the six possibilities, listed in table~\ref{tab:configurations}, that satisfy the close-packing condition, only two are inequivalent up to a redefinition of the lattice vectors. They correspond to the same relative translation between the layers as the HT and HT' discussed above, but with the second layer now rotated by $60^\circ$, and are thus denoted rHT and rHT'. Differently from the case of pure translations when the HT configuration is stable while the HT' is not, the rHT and rHT' stacking sequences both satisfy the close-packing condition. Even more compelling, the two arrangements are energetically indistinguishable as they differ only by the definition of the positive vertical direction. The only symmetry element is a vertical mirror plane that contains the translation vector between the layers, so that such configurations are monoclinic.

\begin{figure*}
\includegraphics{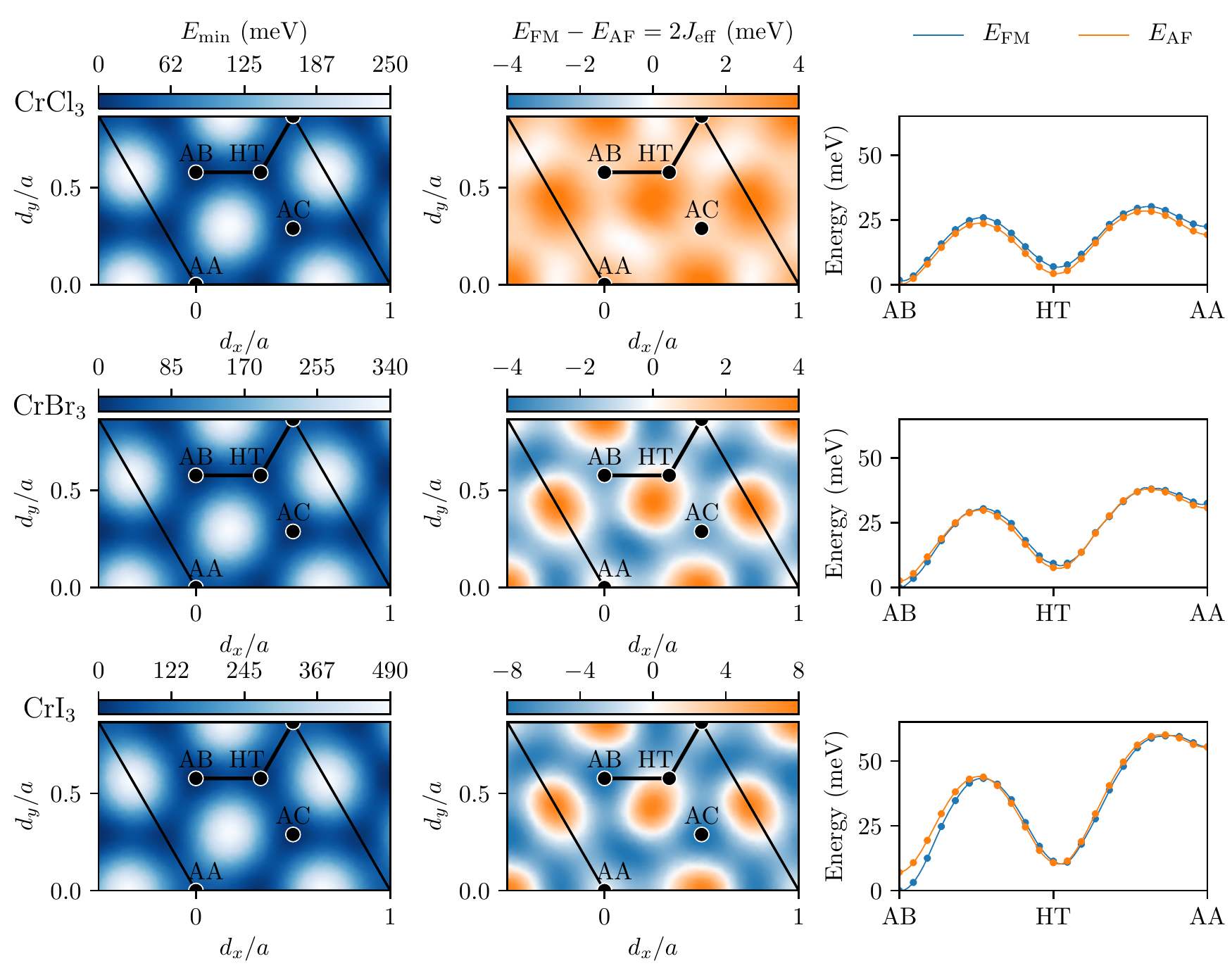}
\caption{Relative stability and interlayer magnetic order in CrX\3 bilayers when both layers share the same chirality. Left: Minimum energy between the ferromagnetic (FM) and antiferromagnetic (AF) configurations, $E_{\rm min} = {\rm min} \{E_{\rm FM},E_{\rm AF}\}$, for bilayer CrX\3 (X = Cl, Br, I, from top to bottom), as a function of the relative in-plane displacement ${\bm d}=(d_x,d_y)$ between the layers at fixed interlayer distance $d_z$. Black circles with a white contour mark the positions corresponding to the stacking patters listed in table~\ref{tab:configurations} that satisfy the close-packing condition. Only one of the three equivalent HT configurations is highlighted. Center:  Similar plot as on the left, but for the energy difference between the FM and AF configuration, related to the effective interlayer exchange coupling $J_{\rm eff}$. Blue regions correspond to a preferential FM order, while orange regions correspond to an AF interlayer coupling. Right: Energy of the FM (blue) and AF (orange) configurations of bilayer CrX\3 along the path highlighted with a thick black line in the left and center panels. For each point along the path the atomic structure, including the interlayer distance, is now optimised while keeping the in-plane positions of Cr atoms fixed. 
\label{fig:scan}}
\end{figure*}

\begin{figure*}
\includegraphics{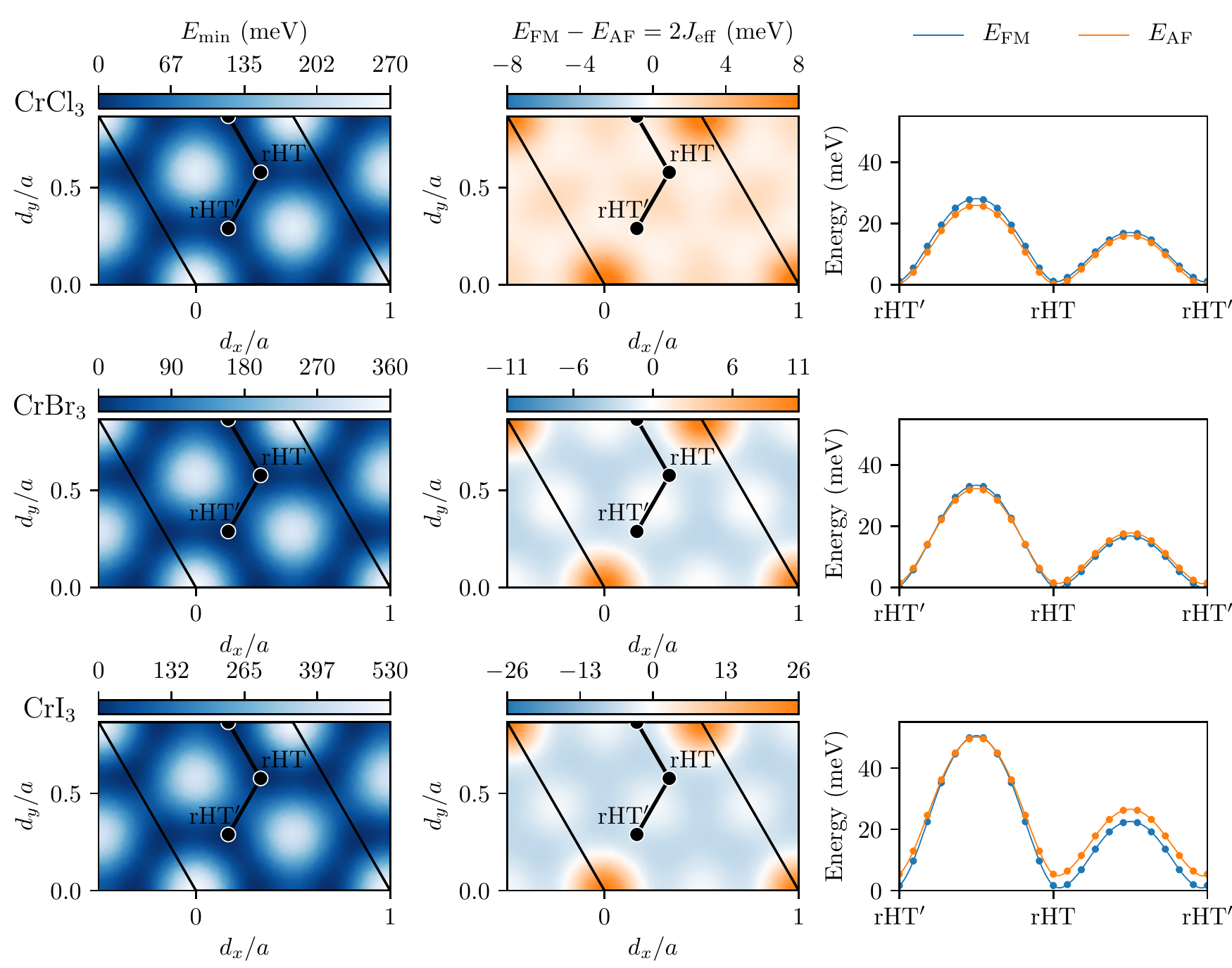}
\caption{Relative stability and interlayer magnetic order in CrX\3 bilayers when the layers have opposite chirality. Left: Minimum energy between the ferromagnetic (FM) and antiferromagnetic (AF) configurations, $E_{\rm min} = {\rm min} \{E_{\rm FM},E_{\rm AF}\}$, for bilayer CrX\3 (X = Cl, Br, I, from top to bottom), as a function of the relative in-plane displacement ${\bm d}=(d_x,d_y)$ between the layers at fixed interlayer distance $d_z$. Black circles with a white contour mark the positions corresponding to the stacking patters listed in table~\ref{tab:configurations} that satisfy the close-packing condition. Only one of the three equivalent rHT and rHT' configurations is highlighted. Center:  Similar plot as on the left, but for the energy difference between the FM and AF configuration, related to the effective interlayer exchange coupling $J_{\rm eff}$. Blue regions correspond to a preferential FM order, while orange regions correspond to an AF interlayer coupling. Right: Energy of the FM (blue) and AF (orange) configurations of bilayer CrX\3 along the path highlighted with a thick black line in the left and center panels. For each point along the path the atomic structure, including the interlayer distance, is now optimised while keeping the in-plane positions of Cr atoms fixed. 
\label{fig:scanrot}}
\end{figure*}

To verify that the configurations listed in table~\ref{tab:configurations}, predicted from crystallographic arguments based on the close-packing condition, are the only (meta)stable primitive stacking sequences for bilayer CrX\3, we perform first-principles DFT simulations as a function of the relative in-plane displacement ${\bm d}=(d_x,d_y)$ between the two layers, with either the same or opposite chirality. The vertical separation $d_z$ between the layers is either kept fixed or optimised as specified below. Calculations are carried out  assuming both a FM and a AF alignment between the layers, with a corresponding total energy $E_{\rm FM}({\bm d})$ and $E_{\rm AF}({\bm d})$. For each displacement ${\bm d}$, it is then possible to evaluate the ground-state energy $E_{\rm min} = {\rm min}\{E_{\rm FM},E_{\rm AF}\}$ and the energy difference $E_{\rm FM}-E_{\rm AF}$ expressing the preference towards a FM/AF alignment. In particular, we can relate the energy difference to an effective interlayer exchange coupling $J_{\rm eff}$, i.e.\ $2 J_{\rm eff} = E_{\rm FM}-E_{\rm AF}$, with $J_{\rm eff} > 0$ leading to an AF state while a FM alignment is expected when $J_{\rm eff} < 0$.

We first consider the case of two layers with the same chirality. The first column of Fig.~\ref{fig:scan} shows the minimum energy $E_{\rm min}$ as a function of $d_x$ and $d_y$ while keeping fixed $d_z$ at the bulk (rhombohedral) value for all CrX\3 bilayers with X = Cl, Br, and I. In all cases, two equivalent global minima are found at a relative displacement corresponding to the AB and AC configurations, and thus to the rhombohedral stacking sequence. This is consistent with the rhombohedral structure being the most stable phase at low temperature. Additional local minima are present and correspond to the HT and AA configurations. In particular, the monoclinic HT arrangement is very close in energy to the stable AB configuration, consistently with the experimental observation of the monoclinic phase at sufficiently high temperature. As expected, the HT' configuration is unstable and represents a local maximum for $E_{\rm min}$. 

In the central panels we show the energy difference between the FM and AF configurations in order to assess the preferential interlayer magnetic order of  a given stacking arrangement. As already pointed out~\cite{Wang2018,Sivadas2018,Soriano2019,Jiang2019theory,Jang2019,Lei2019}, for bilayer CrI\3 there is a strong connection between the stacking sequence and the interlayer spin alignment. The AB arrangement is FM, while the AA and HT configurations have a mild preference for an AF order. A similar situation is found for CrBr\3, although with a tendency to increase $E_{\rm FM}-E_{\rm AF}$ with respect to CrI\3, and thus to suppress FM order in favour of the AF state. The suppression is further enhanced for CrCl\3, where the AF alignment is preferred irrespective of the relative translation between the layers, although $E_{\rm FM}-E_{\rm AF}$ largely depends on $d_x$ and $d_y$. In all cases, the largest AF interlayer exchange coupling is obtained for a configuration corresponding to ${\bm d}=(a/2,a/2)$, dubbed ``special'' in Ref.~\onlinecite{Chen2019}. This is consistent with the AF order measured in Ref.~\cite{Chen2019} for synthetic CrBr\3 bilayers. Nonetheless, it is surprising that this configuration is experimentally accessible since, according to the present simulations, it should not be dynamically stable as it does not represent a local minimum (nor a local maximum) in the energy landscape. 

To compare in more detail the energy of the FM and AF alignments, in the right panels we show their dependence on the stacking sequence along a path (passing through the metastable configurations) highlighted with thick black lines in the left and central panels. In this case, the atomic structure is fully relaxed, including the interlayer distance $d_z$, while keeping fixed the in-plane coordinates of the Cr atoms in order to maintain a given configuration during the force minimisation. All ground-state magnetic orderings discussed above are confirmed, although with a better estimation of the interlayer $J_{\rm eff}$. These plots also give more insight on energy barriers separating the three metastable configurations. In particular,  the very small potential barrier protecting the AA arrangement might be responsible (together with the high energy difference with respect to the AB and HT configurations) for its absence in current experiments under standard conditions, both for bulk and atomically thin samples. 

Interestingly we also note that the energy minimum associated with the HT configuration does not necessarily occur for the expected translations in table~\ref{tab:configurations} (e.g. ${\bm d}=(a,a\sqrt{3})/3$), and might even be slightly different for the FM and AF state. This is due to the fact that in the HT configuration the symmetry is reduced to monoclinic and the positions of Cr atoms (and thus the relative translation) are not enforced by symmetry, although deviations from the ideal structure are typically small.

The above situation changes when the two layers have different chirality, i.e.\ when the second layer is obtained from the first one by performing a $60^\circ$ rotation (or a vertical mirror reflection) before the relative translation. The corresponding energy landscape  is reported in Fig.~\ref{fig:scanrot}. Although stacking configurations are indistinguishable from the ones in Fig.~\ref{fig:scan} when only Cr atoms are considered, the energy profile is completely different, signalling the uttermost importance of the halogen arrangement in determining the stability of a stacking sequence. Moreover, also the difference between the FM and AF alignment is largely affected. This means that extending the current results to arbitrary rotation angles and non-primitive unit cells (relevant for twisted bilayers) is far from trivial, as the precise location of the halogens  --and not only of the Cr atoms-- is crucial in determining the stability and magnetic order of bilayers. 

A more detailed analysis shows that the only local minima in $E_{\rm min}$ correspond to the rHT and rHT' identified in table~\ref{tab:configurations} using crystallographic arguments. Similarly, the local maxima correspond to the rotated analogues of the AA, AB, and AC arrangements (denoted rAA, rAB, and rAC), which are correctly marked as unstable by the close-packing condition as in this case the halogen layers facing the van-der-Waals gap would sit exactly on top of each other. Allowing for atomic relaxation, including the interlayer distance, provides a more accurate estimation of the energy profile between the local minima at the rHT and rHT' configurations, as shown in the right panels for configurations along the path highlighted with thick black lines in the other panels. The two rotated monoclinic sequences are energetically equivalent, as expected, but are separated by different barriers. The highest barrier occurs when passing though a saddle point in between the rAB and rAC configurations, while the barrier is lower when the saddle point is between the rAA and the rAB (or rAC) arrangements. 

Concerning the magnetic order, bilayer CrCl\3 prefers an AF alignment irrespective of the relative translation between the layers also in this case of opposite chirality. CrBr\3 and CrI\3 behave similarly, with a FM ground state favoured for most configurations, including in particular the locally stable rHT and rHT', while AF order would be present only  close to the unstable rAA stacking sequence. Interestingly, we predict a FM alignment in bilayer CrBr\3 for ${\bm d}/a \simeq (0.65,0.17)$, consistently with the  observations in Ref.~\cite{Chen2019}. Nonetheless, also in this case the experimentally reported stacking sequence, dubbed ``bridge I''~\cite{Chen2019}, is predicted to be dynamically unstable.

\begin{table}
\caption {Energy difference (in meV) between the ferromagnetic (FM) and antiferromagnetic (AF) configuration of the (meta)stable stacking patterns of bilayer CrX\3. In parenthesis the minimum energy (in meV) between the FM and AF state is also reported with respect to the stable AB configuration.
\label{tab:summary}}
\vspace{2mm}
\begin{tabularx}{\linewidth}{X C{2.5cm} C{2.5cm} C{2.5cm} }
\toprule
& \multicolumn{3}{c}{{\color{fm} $E_{\rm FM}$} $-$ {\color{af}$E_{\rm AF}$}$=2J_{\rm eff}$~($E_{\rm min}$)}\\
\cmidrule{2-4}
 & $\phantom{-}$CrCl\3 & $\phantom{-}$CrBr\3 & $\phantom{-}$CrI\3 \\
\midrule
AB & {\color{af} $\phantom{-}  1.3$}~($0.0$) &{\color{fm} $-2.8$}~($0.0$) &{\color{fm} $-8.0$}~($0.0$)\\
HT & {\color{af} $\phantom{-}  2.5$}~($3.8$) &{\color{af} $\phantom{-}  1.3$}~($7.8$) &{\color{af} $\phantom{-1}  0.2$}~($10.7$)\\
AA & {\color{af} $\phantom{-1}  3.5$}~($20.0$) &{\color{af} $\phantom{-1}  1.3$}~($32.0$) &{\color{af} $\phantom{-1}  0.3$}~($57.1$)\\
rHT & {\color{af} $\phantom{-}  1.0$}~($5.1$) &{\color{fm} $-1.1$}~($8.8$) &{\color{fm} $\phantom{} -4.6$}~($12.8$)\\
\bottomrule
\end{tabularx}
\end {table}

Having verified that the locally stable primitive stacking configurations are  only the ones expected from the close-packing condition in table~\ref{tab:configurations}, it is of uttermost importance to provide more accurate estimates of the relative stability and effective interlayer exchange coupling for each of them. We thus perform full structural relaxations with tight thresholds for all metastable stacking sequences, allowing also for the lattice parameter and the cell angle (for monoclinic structures) to be optimised. Final results are summarised in table~\ref{tab:summary}. In all cases, the rhombohedral AB stacking sequence is the most stable, followed by  HT and rHT (or equivalently rHT'). The AA arrangement is typically quite high in energy, consistently with its absence in experiments. 

Although numerical values might depend on details of the calculations~\cite{Jang2019,Soriano2020}, general trends can be clearly identified concerning the magnetic ground state of different stacking sequences.  Irrespective of the configuration, AF order is  suppressed --possibly in favour of FM order-- as we go from Cl to Br, to I, with $J_{\rm eff}$ decreasing and even going from  positive to negative. All metastable configurations are AF for CrCl\3, while the AB and rHT arrangements are preferably FM for CrBr\3 and CrI\3, with an effective exchange coupling more negative for the iodide than for the bromide. For all halides, the effective interlayer exchange coupling is largest and positive (i.e.\ AF) for the AA configuration, followed by the HT sequence, and further decreases for the AB and rHT arrangements (possibly becoming negative for Br- and I-based bilayers).

These trends are consistent with all current experimental observations and have strong implications for future measurements. The AB configuration has a negative $J_{\rm eff}$ for Br- and I- based bilayers, while it is positive for CrCl\3, consistent with the  magnetic ordering measured at low temperature in bulk rhombohedral structures. The interlayer alignment becomes AF for CrI\3 when considering the HT configuration, in agreement with the AF state observed in atomically thin samples~\cite{Song2018,Klein2018,Wang2018,Kim2018} with a monoclinic stacking sequence~\cite{Ubrig2019,Sun2019}. Even more compelling, the increase in the AF $J_{\rm eff}$ for CrCl\3 when comparing the AB (rhombohedral) and HT (monoclinic) configurations is perfectly compatible with recent experimental observations of an enhanced interlayer exchange coupling in  thin (monoclinic) samples with respect to bulk (rhombohedral) crystals~\cite{Klein2019}. Finally, although  no experimental information is available on the stacking sequence of CrBr\3 bilayers, the current results suggest that the measured FM order~\cite{Ghazaryan2018,Kim2019hall,Kim2019} would be compatible only with a AB rhombohedral configuration. This prediction is consistent with the fact that CrBr\3 is in the rhombohedral phase at the temperatures at which exfoliation takes place, and the same AB arrangement is inherited by the exfoliated multilayers. On the contrary CrI\3 and CrCl\3 are monoclinic when exfoliated, and apparently thin samples are not able to undergo the bulk structural phase transition~\cite{Ubrig2019,Klein2019,Sun2019} and remain monoclinic also at the low temperatures at which magnetism sets in.

\section{Conclusions}

We consider chromium trihalide bilayers and investigate all possible stacking sequences that preserve the translational symmetry of each layer. We first identify a set of configurations based on crystallographic arguments by imposing  a close-packing condition. This analysis not only recovers  the stacking patterns observed in bulk structures but also predicts configurations with no bulk counterpart where the two layers have different \emph{chirality}, which have been recently observed in synthetic bilayers. By performing first-principles simulations we validate that these configurations are the only (meta)stable primitive stacking sequences and we associate to each of them a preferential interlayer magnetic ordering. These predictions are consistent with (and provide an explanation for) all current experiments on bulk and exfoliated CrX\3 crystals, ranging from the different magnetic order in thin and bulk CrI\3 to the enhanced antiferromagnetic interlayer exchange coupling in CrCl\3 bilayers. For CrBr\3, our simulations suggest that atomically thin samples should display a rhombohedral stacking sequence to account for the observed FM order, consistently with the fact that bulk crystals are already in the rhombohedral phase when exfoliated. Our results are also compatible with the magnetic ordering observed in non-standard stacking sequences observed in CrBr\3 bilayers grown by molecular beam epitaxy, although we predict that such configurations should not be dynamically stable. Finally, the dramatic differences  between bilayers where the two layer have the same or opposite chirality clearly shows the importance of the precise arrangement of also the halogen atoms in determining the stability and magnetic ground state of a given configuration, so that caution should be used in extending the current results to arbitrary stacking sequences and twist angles.

\section{Acknowledgements}
Alberto Morpurgo, Zhe Wang, Nicolas Ubrig, and Ignacio Guti\'errez-Lezama are greatly acknowledged for useful discussions.
Support has been provided  by the Italian Ministry for University and Research through the Levi-Montalcini program. I acknowledge support during the early stage of this project also by the Swiss National Science Foundation (SNSF) through the Ambizione program (grant PZ00P2\_174056) and by the NCCR MARVEL funded by the SNSF. Simulation time was provided by CSCS on Piz Daint (production project s917). 

\bibliographystyle{iopart-num}
\bibliography{CrX3_bilayers}

\end{document}